\documentclass[review,1p]{elsarticle}
\usepackage{graphics}
\usepackage[ruled]{algorithm2e}
\usepackage{graphicx}
\usepackage{epsfig}

\usepackage{amssymb,amsmath,amsfonts}
\usepackage{amsthm}

\newcommand{\xb}{\mathbf{x}}

\newcommand{\yb}{\mathbf{y}}

\newcommand{\Qb}{\mathbf{Q}}

\newcommand{\ub}{\mathbf{u}}

\newcommand{\wb}{\mathbf{w}}
\newcommand{\muB}{\mbox{\boldmath$\mu$}}

\journal{Mathematical Biosciences}
\begin{document}

\begin{frontmatter}


\title{Monitoring and prediction of an epidemic outbreak using syndromic observations}


\author{Alex Skvortsov and Branko Ristic }

\address{Defence Science and Technology Organisation\\
506 Lorimer Street, Melbourne, VIC 3207\\Australia}

\begin{abstract}
%Syndromic surveillance has become an important means for detection,
%prediction and mitigation of emerging epidemics.
The paper presents
an algorithm for syndromic surveillance of an epidemic outbreak
formulated in the context of stochastic nonlinear filtering.  The
dynamics of the epidemic is modeled using a generalized
compartmental epidemiological model with inhomogeneous mixing. The
syndromic (typically non-medical) observations of the number of
infected people (e.g. visits to pharmacies, sale of certain
products, absenteeism from work/study etc.) are used for estimation.
The state of the epidemic, including the number of infected people
and the unknown parameters of the model, are estimated via a
particle filter. The numerical results indicate that the proposed
framework can provide useful early prediction of the epidemic peak
if the uncertainty in prior knowledge of model parameters is not
excessive.
\end{abstract}

\begin{keyword}
Emerging epidemics, bio-terrorism, inhomogeneous mixing, social
network, mathematical biology, nonlinear filtering, particle filter,
imprecise likelihood.
\end{keyword}

\end{frontmatter}

\renewcommand{\baselinestretch}{1.5}
\section{Introduction}
\label{s:intro}

\noindent Epidemics can impose significant challenges on modern
societies, not only by affecting the health of the general
population, but also by causing negative trends in the economy
(medical treatments, absenteeism from work,  missed business
opportunities, etc). The ongoing epidemics of AIDS, tuberculosis and
the recent outbreaks of SARS and H1N1 (swine flu) provide some
revealing examples. In the absence of an effective cure against many
diseases, the best approach to mitigate an epidemic outbreak
(malicious or natural) resides in the development of capability for
its early detection and for prediction of its further development
\cite{Wagner_06}, \cite{Wilson_06}, \cite{Mnatsakanyan_10}, \cite{Cauchemez}, \cite{Dailey}, \cite{Fraser}.  Such a capability
would allow making any countermeasures (quarantine, vaccination,
medical treatment) much more effective and less costly
\cite{Walden}, \cite{Wilson_06}.

Syndromic surveillance is refereed to as systematic collection, analysis,
and interpretation of public health data for the purpose of early (and often cost effective)
detection of an epidemic outbreak, its continuous monitoring, and
timely response by public health agencies
\cite{Wagner_06}, \cite{Wilson_06}, \cite{chenzeng_2010}. The rationale for this method rests on a resalable assumption  that a spread of an
infectious disease is usually associated with the measurable changes
in the social behaviour.

It is important to highlight that although the measurements (data
streams) for syndromic surveillance often rely on the medical
observations (the visits to medical practitioners, occurrence of
particular symptoms, the number of diagnosed cases, etc) they are
not restricted to the pure medical data. Recent studies
\cite{AMIA2006}, \cite{Schuster_10}, \cite{KnDiscDataMining}, have
demonstrated that  ``non-medical'' sources of syndromic data
streams, such as the absenteeism from work/school, the
pharmaceutical sales, Internet queries, Twitter messages, and alike,
can enable one to draw important conclusions regarding the epidemic
state in the community. The `Google Flu' project \cite{Ginsberg_09}
(flu-related searches in Google), is a well publicised example of
this approach.
%Such a diversification of data streams
% provides flexible options for the scalability of associated data processing
% algorithms and for feasibility of their practical implementation.

The algorithms for syndromic surveillance and its practical
implementation have recently attracted significant attention by
scientists and practitioners; there is a vast amount of literature
devoted to this topic (for more comprehensive review see
\cite{Wagner_06}, \cite{Wilson_06} and refs therein).  In general,
all algorithms applied in this area can be divided into two main
groups, the \emph{data mining} methods and the \emph{information
fusion} (also known as {data assimilation}) methods. Data mining is
primarily concerned with the extraction of patterns from massive
amounts of raw data without using dynamic models of the underlying
process (i.e. epidemic spread) \cite{Ginsberg_09},
\cite{KnDiscDataMining}. Information fusion algorithms, on the
contrary, strongly rely of mathematical models: in this case, the
{\em dynamic model} of an epidemic outbreak and the {\em measurement
model} of a particular syndromic data stream  \cite{Egat_08},
\cite{Cazelles_97} , \cite{Mandela_10}. Evidently  the accuracy of
the information fusion algorithms is significantly determined by the
fidelity of the underlying models.

%There is no conclusive decision regarding comparative  effectiveness
%of these group of algorithms in application to the problem of
%syndromic surveillance. The choice of an algorithm mostly depends on
%the operational context and, in general,  these algorithms are
%complementary. For instance, the  data mining methods  (which seems
%to be more popular at the moment) are evidently more flexible (since
%they are not constrained by any underlying dynamical model), but
%they usually require massive amount of data. Contrarily, the data
%assimilation methods are more effective with a limited dataset, but
%they, indeed, are constrained by the ''fidelity'' of the postulated
%dynamical model of epidemic.

This paper presents a study of a recursive information fusion
algorithm for syndromic surveillance. The problem is formulated in
the Bayesian context of stochastic nonlinear filtering and  solved
using a particle filter. While this problem formulation and the
particle filter implementation have been considered earlier, see
\cite{jegat_epid}, \cite{icassp09}, \cite{ong_10},
\cite{skvortov_10_fus}, this paper introduces several novelties. In
order to overcome  the limitations  of the standard ``compartment''
model of epidemic (the ``well-mixed'' approximation) we employ a
more flexible alternative, initially proposed in \cite{stroud_06}
(and later extended in \cite{Novozhilov}, \cite{Roy}). The adopted
epidemic model has the explicit parameter of ``mixing efficiency''
(or level of social interaction) and is therefore more appropriate
to represent the variety of social interactions in a small community
(self-isolation and panic). A significant advantage of the adopted
epidemiological model is a rigorous estimation of its noise
component (scaling law of noise level  with the population size of a
community, see below) resulting in more accurate parameter
estimations.  Furthermore, a more flexible model of syndromic
measurements, validated with a limited data set available in the
literature  \cite{Ginsberg_09}, \cite{AMIA2006}, is adopted in the
paper. This measurement model is derived from \cite{Ginsberg_09},
which has been extensively validated by observations and is allowed
to have imprecisely known parameters. The optimal sequential
estimator (filter) and predictor are then formulated in the Bayesian
framework and solved using a particle filter.  The paper  includes
numerical results that verify the theoretical framework and analyses
the sensitivity to partially known epidemic parameters. From this
perspective, the present study is a significant extension of our
earlier work \cite{icassp09}, \cite{skvortov_10_fus}.

\section{Problem specification}
\label{sec:mod}

This section describes the models to be used in estimation.  To
describe the dynamics of an epidemic outbreak we employ the
generalized  SIR epidemic model proposed in \cite{dargatz},
\cite{Dangerfield}, \cite{anderson_04}, \cite{Daley_book} with
stochastic fluctuations. According to this model the population of
the community can be subdivided into three interacting groups:
susceptible, infectious and recovered individuals. Let the number of
susceptible, infectious and recovered be denoted by $S$, $I$ and
$R$, respectively, so that $S+I+R=P$, where $P$ is the total
population size. The dynamic model of epidemic progression in time
can be be expressed by two stochastic differential equations
\cite{anderson_79}, \cite{dargatz}, \cite{Dangerfield} and the
``conservation'' law for the population as follows:
\begin{eqnarray}
\frac{ds} {dt}  &=& - \alpha\; i\; s^\nu +  \sigma_q \xi, \label{e:sir1}\\
\frac{di} {dt}  & = & \alpha\; i\; s^\nu  - \beta i  -  \sigma_q\xi + \sigma_{\beta}\zeta,  \label{e:sir2}\\
           r & = & 1 - s - i. \label{e:sir3}
\end{eqnarray}
The explanation of notation: $s = S/P$, $i = I/P$,  $r = R/P$ and
$\xi,\zeta $ are two uncorrelated white Gaussian noise processes,
both with zero mean and unity variance. The terms $\sigma_q \equiv
\sigma_q (s, i)$ and $ \sigma_{\beta} \equiv \sigma_{\beta} (s, i) $
are introduced to capture the demographic noise (random variations
in the contact rate $\alpha$ and in the recovery time $\beta$
between individuals), for details see \cite{Herwaarden},
\cite{dargatz}, \cite{Dangerfield}.

The parameters of this model are described next: $\beta$ is the
recovery time, that is a reciprocal of the average infectious period
for the disease; $\alpha = \rho\cdot \beta$, where $\rho$ is a well
known parameter in epidemiology, referred to as the {\em the basic
reproductive number}, which represents the average number of people
infected by a direct contact with a sick person. Finally $\nu$ is
the population mixing parameter, which for a homogeneous population
equals $1$. In the presence of an epidemic, $\nu$ may vary as people
may behave unpredictably (panic) or avoid the risk of infection
(self-isolation). In general, the epidemic model parameters can be
assumed to be partially known as interval values, that is $\alpha\in
[\underline{\alpha}, \overline{\alpha}]$, $\beta\in
[\underline{\beta}, \overline{\beta}]$ and $\nu\in [\underline{\nu},
\overline{\nu}]$.

The amplitude of the noise terms $ \sigma_q, \sigma_{\beta}$  in
(\ref{e:sir1}) and (\ref{e:sir2}) can be established from a scaling
law of Gaussian fluctuations generated by the random contact rate
$q=\alpha\,i\,s^\nu$ and recovery time $\beta$. Thus for a dynamical
system (\ref{e:sir1}) - (\ref{e:sir3}) consisting of a large number
of individuals $P$ we can write (for details see \cite{dargatz},
\cite{Dangerfield}, \cite{Herwaarden})

\begin{equation}\label{eqn:E60}
                \sigma_q(s, i) \approx
                \sqrt{\frac{{\alpha\; i\; s^\nu}}{P}}
                ,~~
                \sigma_{\beta}(s, i)  \approx
                \sqrt{\frac{{\beta} i}{P}}.\\
\end{equation}

With further reasonable assumptions $s \approx  s_0 \approx 1$, $r \approx  r_0 \approx 0$, $i \approx  i_0 \approx  1/P$, that holds for the initial stage of epidemic (where $s_0$, $i_0$, $r_0$ are initial values of $s, i, r$)  expressions (\ref{eqn:E60}) can be reduced to the following scaling  for the noise components of the model
\begin{equation}\label{eqn:E6}
                \sigma_q(s, i) \approx
                \frac{\sqrt{\alpha}}{P}
                ,~~
                \sigma_{\beta}(s, i)  \approx
                \frac{\sqrt{\beta}}{P}.\\
\end{equation}

The measurement model is introduced next. Following
\cite{Ginsberg_09}, \cite{KnDiscDataMining}  we can assume that
a power law relationship holds for odds-ratios of syndromic
observations and number of infected people:
\begin{equation}
\label{eqn:E20}
    \frac{z_{j}}{1 - z_{j}}\propto \left ( \frac{i}{1 - i} \right)^{\varsigma_j},
\end{equation}
where $z_{j}$ is the observable syndrome with index $j=1,\dots,N_z$
(normalized by  population size $P$),  $\varsigma_j$ is the power-law exponent (in general
different for different syndromes). Since  at the initial stage of
epidemic $i \ll 1, ~ z_{j} \ll 1$, eq.(\ref{eqn:E20}) can be reduced
to a simple power-law model
\begin{equation}
\label{eqn:E21}
     z_{j} = b_j\, i^{\varsigma_j} + \tau_j,
\end{equation}
where $b_j = const$. The noise term $\tau_j $ is added  to simulate
random nature of measurement (measurements noise) which is assumed
to be uncorrelated to other syndromes and noises $\xi $ and $\zeta$.
Since  $z_{j} \ge 0$ the noise  term  $\tau_j$  associated with
syndrome $j$ should be modelled with a random variable that provides
strictly non-negative realisations (e.g. Poissonian
\cite{Mandela_10} or truncated Gaussian \cite{dargatz}). In the
current study we use a random variable that obeys the log-normal
distribution: this means that we set $\tau_j = \sigma_j \eta_j$,
where $\eta_j$ is the standard log-normal noise  $\eta_j \sim
\ln{\cal N}(0,1)$, see below.

Parameters $b_j,  \sigma_j, \varsigma_j$  typically are not known,
but with a representative dataset of observations the model
(\ref{eqn:E21}) can be easily calibrated (see results of the linear
regression fits in \cite{Ginsberg_09}). The data fit reported in
\cite{AMIA2006} suggests that $\varsigma_j$ may be close to unity
(but it is not precisely known because of significant scattering of
data points). To cater for this uncertainty  we assume that
$\varsigma_j$ can take any value in an interval, $\varsigma \in
[\underline{\varsigma},\overline{\varsigma}]$ around $\varsigma =1$
(for the sake of simplicity we also assume that all $\varsigma_j =
\varsigma = const$). Unfortunately  \cite{Ginsberg_09},
\cite{KnDiscDataMining},  do not report any specific values of
fitting parameters, so we have to use some heuristic values for
$b_j, \sigma_j$ in our simulations.

The scaling law for the noise parameter $\sigma_j$ in
(\ref{eqn:E21}) can be deduced by employing the arguments leading to
(\ref{eqn:E6}) and thus we arrive at the different scaling
$\sigma_j \propto (1/P)^\epsilon$, $\epsilon = \min \{(\varsigma
+1)/2, 1/2 \}$. The later scaling law in conjunction with the
scaling laws (\ref{eqn:E6}) provides a consistent way to compare predictive skills of our algorithm for communities
with different population size and to draw  conclusive decisions
about its performance before any operational deployment.

%
%\begin{figure}[ht0]
%\centerline{\includegraphics[height=4cm,width=7.5cm]{Z_I.eps}}
% \caption{The number of the flu-related web searches as a function of the number of flu-diagnosed cases in the community. The dotted line is a linear fit corresponding to the model (\ref{eqn:E21}) with $\varsigma =1$.}
% \label{f:mes}
%\end{figure}
%

 Being formulated in the Bayesian framework, our problem is to estimate the
(normalised) number of infected $i$ and susceptible $s$ at time $t$,
using syndromic observations  $z_{j}$  (\ref{eqn:E21}) collected up to time $t$. Let $\xb$
denote the state vector to be estimated; it includes $i$ and $s$,
but also the imprecisely known parameters $\alpha$, $\beta$ and
$\nu$. The formal Bayesian solution is given in the form of the
posterior probability density function (PDF) $p(\xb_t|z_{1:t})$,
where $\xb_t$ is the state vector at time $t$ and $z_{1:t}$ denotes
all observations up to time $t$. Using the posterior
$p(\xb_t|z_{1:t})$ one can predict the progress of the epidemic
using the dynamic model (\ref{e:sir1})-(\ref{e:sir3}).

\section{Optimal Bayesian solution for imprecise likelihood}

\subsection{Formulation}

The model (\ref{e:sir1})-(\ref{e:sir3}) is given in continuous
time. For the purpose of computer implementation, we require a
discrete-time approximation of this model. The state vector is
adopted as
\begin{equation}
\xb = \left[\begin{matrix}i&s& \alpha & \beta &
\nu\end{matrix}\right]^\intercal
\end{equation}
where $^\intercal$ denotes matrix transpose. Neglecting for the
moment the process noise terms, the evolution of the epidemic state
can be written according to (\ref{e:sir1})-(\ref{e:sir2}) as
$\dot{\mathbf{x}}=\mathbf{g}(\mathbf{x})$ where
$\mathbf{g}(\mathbf{x}) = \left[\begin{matrix}(\alpha s^\nu -\beta)i
& -\alpha i s^\nu & & 0 &0 &0\end{matrix}\right]^{\intercal}$. The
nonlinear differential equation governing the evolution of the state
cannot be solved in the closed-form. The Euler method provides a
simple approximation valid for small integration interval $\tau>0$:
$ \mathbf{x}(t+\tau) \approx \mathbf{x}(t) +
\tau\mathbf{g}(\mathbf{x}(t))$. The state-evolution in discrete-time
$t_k$ can then be expressed as:
\begin{equation}
\mathbf{x}_{k+1} \approx \mathbf{f}_k(\mathbf{x}_k) + \wb_k
\label{e:dyn}
\end{equation}
where $k=t_k/\tau$ is the discrete-time index and
$\mathbf{f}_k(\mathbf{x}_k)$  is the transition function  given by
\begin{equation}
\mathbf{f}_k(\xb_k) = \left[ \begin{matrix} \xb_k[1] +
\tau\xb_k[1]\left(\xb_k[3]\xb_k[2]^{\xb_k[5]}-\xb_k[4]\right)
\\ \xb_k[2] -\tau \xb_k[3]\xb_k[1]\xb_k[2]^{\xb_k[5]} \\
\xb_k[3] \\
\xb_k[4] \\
\xb_k[5]
\end{matrix}\right] \label{e:f}
\end{equation}
In this notation $\xb_k[i]$ represents the $i$th component of vector
$\xb_k$. Process noise $\wb_k$ in (\ref{e:dyn}) is assumed to be
zero-mean white Gaussian with a diagonal covariance matrix $\Qb$,
which according to (\ref{eqn:E6}) can be expressed as $\Qb \approx
\mbox{diag}[(\alpha+\beta)\tau^2/P^2,\;\; \alpha\tau^2/P^2]$.

The optimal Bayes filter  is typically presented in two steps, {\em
prediction} and {\em update}. Suppose the posterior PDF at time
$t_{k}$ is given by $p(\xb_{k}|z_{1:k})$. Then the prediction step
computes the PDF predicted to time $t_{m} = t_{k} + \tau$ as
\cite{jazwinski_70}:
\begin{equation}
p(\xb_{m}|z_{1:k}) = \int \pi(\xb_{m}|\xb_{k})\;
p(\xb_{k}|z_{1:k})\, d\xb_{k}
\end{equation}
where $\pi(\xb|\xb')$ is the transitional density. Let ${\cal
N}(\yb;\muB,\mathbf{P})$ denote a Gaussian PDF with mean $\muB$ and
covariance $\mathbf{P}$.  Then  according to (\ref{e:dyn}) the
transitional density is given by
\begin{equation} \pi(\xb|\xb') =
{\cal N}(\xb; \mathbf{f}_k(\xb'), \Qb).  \label{e:trden}
\end{equation}
The
prediction step is carried out many times with tiny sampling
intervals $\tau$ until an observation $z_{j,k+1}$ becomes available
about syndrome $j$ at time $t_{k+1}$.

The predicted PDF at $t_{k+1}$ is $p(\xb_{k+1}|z_{1:k})$. In the
standard Bayesian estimation framework, this PDF is updated using
the measurement $z_{j,k+1}$ by multiplication with the measurement
likelihood function \cite{jazwinski_70}. According to
(\ref{eqn:E21}), the likelihood function in this case is
$\ell(z_{j,k+1}|\xb_{k+1}) = \ln {\cal N}(z; h(\xb_{k+1};\varsigma),
\sigma_j^2)$, where $h(\xb_k;\varsigma) = b_j\cdot
\xb_k[1]^{\varsigma}$. Now, the problem is that $h(\xb_k;\varsigma)$
defined is this way is not a function because $\varsigma \in
[\underline{\varsigma},\overline{\varsigma}]$.

An elegant solution to the imprecise measurement transformation is
available in the framework of random set theory \cite{mahler_07}. In
this approach $h(\xb;\varsigma)$ defines a closed set $\Sigma_{\xb}$
on the measurement space $\mathcal{Z}$. The closed set
$\Sigma_{\xb}$ is random because the state $\xb$ is random. In fact,
a random closed set (RCS) $\Sigma_{\xb}$ can be seen as a composite
mapping. The first is the measurable mapping which defines the
random variable $\xb: \Omega\rightarrow \mathcal{X}$, where $\Omega$
is the sample space and $\mathcal{X}$ is the state space. The second
mapping is $h(\xb;\varsigma): \mathcal{X} \rightarrow \mathcal{IZ}$,
where $\mathcal{IZ}$ is the set of closed sets of $\mathcal{Z}$. The
RCS $\Sigma_{\xb}$ is therefore a random variable that takes values
as closed intervals of $\mathcal{Z}$.

The Bayes updated step using measurement $z_{j,k+1}$ is now defined
as \cite{mahler_07}:
\begin{equation}
p(\xb_{k+1}|z_{1:k+1}) =
\frac{\tilde{\ell}(z_{j,k+1}|\xb_{k+1})\cdot
p(\xb_{k+1}|z_{1:k})}{\int \tilde{\ell}(z_{j,k+1}|\xb_{k+1})\cdot
p(\xb_{k+1}|z_{1:k})\, d\xb_{k+1}}
\end{equation}
where $\tilde{\ell}(z_{j,k}|\xb_k)$ is referred to as the
generalised likelihood function. For the measurement model
(\ref{eqn:E21}), $\tilde{\ell}(z_{j,k}|\xb_k)$ is  defined as
\cite{ristic_11}:
\begin{equation}
\tilde{\ell}(z_{j,k}|\xb_k) = \phi(z_{j,k};
\underline{\Sigma}_{\xb_k},\sigma_j^2) - \phi(z_{j,k};
\overline{\Sigma}_{\xb_k},\sigma_j^2)  \label{e:gen_lik_norm}
\end{equation}
where
\begin{eqnarray}
\underline{\Sigma}_{\xb} & = & \min\{h(\xb;\underline{\varsigma}), h(\xb;\overline{\varsigma})\} \label{e:lim1}\\
\overline{\Sigma}_{\xb}  & = & \max\{h(\xb;\underline{\varsigma}),
h(\xb;\overline{\varsigma}) \}. \label{e:lim2}
\end{eqnarray}
and $\phi(\ub;\muB,\mathbf{P})=\int_{-\infty}^\ub \ln{\cal
N}(\yb;\muB,\mathbf{P})\, d\yb$ is the cumulative log-normal
distribution.

The recursions of the  Bayes filter start with the initial PDF (at
time $t_k=0$), denoted $p(\xb_0)$, which is assumed known.

\subsection{Particle filter implementation}

The proposed Bayesian estimator cannot be solved in the closed form.
Instead we develop an approximate solution based on the particle
filter (PF) \cite{pf_tute}, \cite{pfbook}. The PF approximates the
posterior PDF $p(\xb_k|z_{1:k})$ by a weighted random sample, that
is:
\begin{equation}
p(\xb_k|z_{1:k}) \approx \sum_{i=1}^S w_k^{(i)} \,
\delta_{\xb^{(i)}_{k}}(\xb) . \label{e:pf1}
\end{equation}
Here $\delta_{\yb}(\xb)$ is the Dirac delta function focused at
$\yb$, $\xb_{k}^{(i)}$ is a sample (particle), $w_k^{(i)}$ is its
weight and $S$ is the particle count. As $S$ is increased,
approximation (\ref{e:pf1}) becomes more accurate. The weights are
normalised so that:
\begin{equation}
\int p(\xb_k|z_{1:k})\, d\xb_k \approx \sum_{i=1}^S w_k^{(i)} = 1.
\end{equation}

The adopted PF implementation is known as the bootstrap PF
\cite{gordon_etal_93}. The algorithm starts by forming a random
sample from the initial PDF $p(\xb_0)$: $\xb_0^{(i)} \sim p(\xb_0)$,
for $i=1,\dots,S$. The initial weights are equal, that is
$w_0^{(i)}=1/S$. The prediction steps consist of propagating
particles using the dynamic model (\ref{e:dyn}). This is implemented
by drawing samples from the transitional density (\ref{e:trden}).
The prediction steps are carried out until a measurement $z_{j,k}$
becomes available, that is until time $t_k$. Suppose the set of
predicted particles at time $t_k$ is $\{\xb^{(i)}_{k|k-1};
i=1,\dots,S\}$.

The update step is implemented in two stages. First the unnormalised
importance weights of predicted particles are computed as
\cite{pfbook}:
\begin{equation}
\tilde{w}_k^{(i)} = \tilde{\ell}(z_{j,k}|\xb^{(i)}_{k|k-1})
\end{equation}
with generalised likelihood $\tilde{\ell}$ specified in
(\ref{e:gen_lik_norm}). This is followed by the weight
normalisation, that is
\begin{equation}
w_k^{(i)} = \tilde{w}_k^{(i)} /\; \sum_{n=1}^S \tilde{w}_k^{(n)}
\end{equation}
for $i=1,\dots,S$.

The second stage is the resampling step. The role of resampling is
to eliminate (in the probabilistic manner) the particles with low
importance  weights and to clone the samples with high importance
weights. This is carried out by sampling with replacement, with the
probability of sampling each $\xb^{(i)}_{k|k-1}$ equal to the
normalised importance weight $w_k^{(i)}$. The result is a mapping of
a random measure $\{w^{(i)}_{k},\xb^{(i)}_{k|k-1}\}_{i=1}^S$ into a
new random measure with uniform weights:
$\{\frac{1}{S},\xb^{(i)}_{k}\}_{i=1}^S$. Several computationally
efficient resampling schemes have been reported; we have implemented
the {\em systematic resampling} algorithm \cite{kita96b}, following
the pseudo-code given in Table 3.2 of \cite{pfbook}.

\section{Numerical Results}
\label{e:Numerics}
\subsection{Experimental data set}
\begin{figure}[htb]
\centerline{\includegraphics[height=6cm,width=7.5cm]{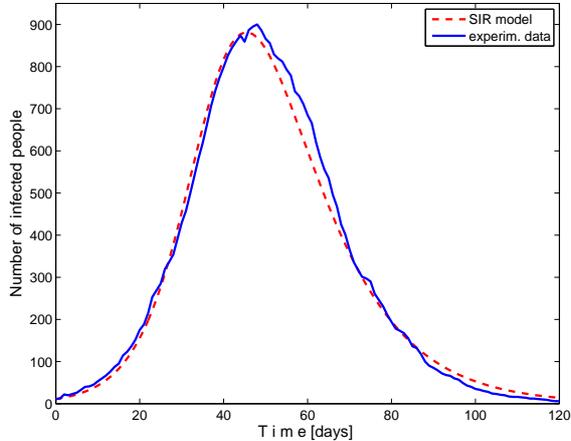}}
 \caption{The solid line, representing the number of infected people over time, was
 obtained from the agent based simulation population model. The
 dashed line shows the generalised SIR model fit}
 \label{f:exper}
\end{figure}

The estimation and prediction of an epidemic will be carried out
using an experimental data set obtained using a large-scale agent
based simulated population model \cite{modsim07}, \cite{biocomp_07}
of a virtual town of $P=5000$ inhabitants, created in accordance
with the Australian Census Bureau data. The agent based model is
rather complex and takes a long time to run. It includes typical
age/gender breakdown and family-household-workplace habits with the
realistic day-to-day people contacts for a disease spread. The blue
line in Fig.\ref{f:exper} shows the number of people of this town
infected by a fictitious disease, reported once per day during a
period of $154$ days (only first $120$ days shown). The dashed red
line represents the deterministic SIR model fit (using the entire
batch of $154$ data points, and setting $\wb_k=0$ in (\ref{e:dyn})),
with estimated parameters $\hat{\alpha}=0.2399$, $\hat{\beta} =
0.1066$, $\hat{\nu}=1.2042$. The parameter estimates were obtained
using importance sampling via the progressive correction algorithm
\cite{OUDJANEMUSSO}. Fig.\ref{f:exper} serves to verify that the
generalized SIR model, although very simple and fast to run, is
remarkably accurate in explaining the data obtained from a very
complex simulation system (for further details see
\cite{stroud_06}).

\subsection{Estimation and prediction results}

The {\em true} number of infected people in simulations was chosen
to be the output of the agent based simulated population model,
shown by the solid line in Fig.\ref{f:exper}. The measurements are
generated synthetically in accordance with (\ref{eqn:E21}), using
the following parameters: $\varsigma = 1.05$ and $b_j=0.25$,
$\sigma_j=0.01$, for all $j=1,2,3,4$ monitored syndromes.
Independent measurements concerning all $N_z=4$ syndromes were
available on a daily basis during the first 25 days. The problem was
to perform estimation sequentially as the measurements become
available until the day number 25, and after that to predict the
epidemic size.

The initial PDF for the state vector was chosen as $p(\xb_0) =
p(i_0)p(s_0)p(\alpha_0)p(\beta_0)p(\nu_0)$ with $p(i_0)  =
\mathcal{N}_{[0,1]}(i_0;z_{j,1}/b_j,\sigma_j^2)$, $p(s_0)  =
\mathcal{N}_{[0,1]}(s_0;1-z_{j,1}/b_j,\sigma_j^2)$, $p(\alpha_0)  =
\mathcal{U}_{[0.18, 0.50]}(\alpha_0)$, $p(\beta_0)  =
\mathcal{U}_{[0.1, 0.143]}(\beta_0)$,  $p(\nu_0)  =
\mathcal{U}_{[0.95, 1.3]}(\nu_0)$, where $\mathcal{N}_{[a,
b]}(x,\mu,P)$  and $\mathcal{U}_{[a,b]}(x)$ denote a truncated
Gaussian distribution restricted at support $[a,b]$ and a uniform
distribution with limits $[a,b]$, respectively. The number of
particles was set to $S=10000$.

\begin{figure}[tbhp]
\centerline{\includegraphics[height=6cm]{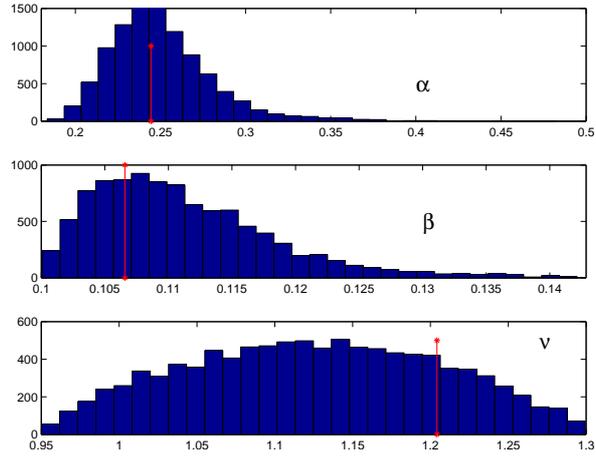}}
\centerline{(a)}
\centerline{\includegraphics[height=6.5cm]{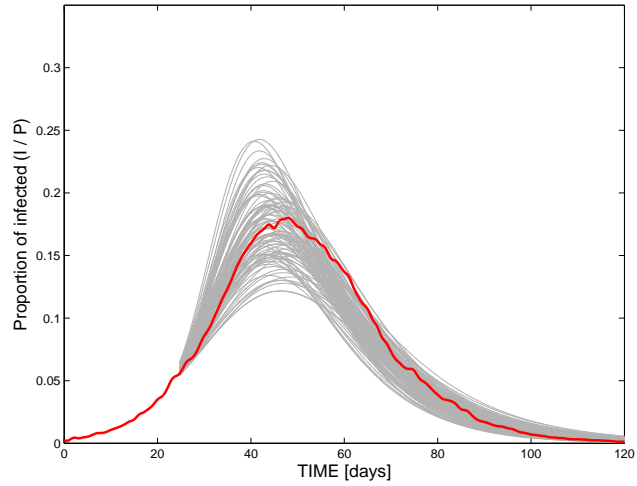}}
\centerline{(b)}
 \caption{Estimation/prediction results from the particle filter after processing measurements collected
 over $25$ days of
surveillance assuming that $\varsigma\in[1.03,1.07]$ (true value
$\varsigma=1.05$): (a) the histograms of estimated parameters
$\alpha$, $\beta$ and $\nu$ (true values indicated by vertical red
lines); (b) Prediction results for a random sample of $100$
particles (gray lines); the red line is the experimental curve from
Fig.\ref{f:exper}}
 \label{f:1}
\end{figure}

The first run of the proposed simulation setup was carried out
assuming that $\varsigma\in[1.03, 1.07]$. This corresponds to the
case of fairly precise knowledge of $\varsigma$ (recall the true
value was $1.05$). Fig. \ref{f:1}.(a) shows the histograms of
particle filter estimated values of $\alpha$, $\beta$ and $\nu$,
after processing all 25 days of measurements (i.e. in total 100
measurements, since $N_z=4$). This figure reveals that the
uncertainty in parameters $\alpha$ and $\beta$ has been
substantially reduced, compared with initial $p(\alpha_0)  =
\mathcal{U}_{[0.18, 0.50]}(\alpha_0)$ and $p(\beta_0)  =
\mathcal{U}_{[0.1, 0.143]}(\beta_0)$. The uncertainty in $\nu$, on
the other hand, has not been reduced, indicating that this parameter
cannot be estimated. While this is unfortunate, it does not appear
to be a serious problem since the prior on $\nu$ in practice is
tight ($\nu\approx 1$). This is confirmed in Fig.\ref{f:1}.(b) which
shows a sample of $100$ overlayed predicted epidemic curves (gray
lines) based on the state estimate after 25 days. Fig.\ref{f:1}.(b)
indicates the prediction performance; the timing of the peak of the
epidemic appears to be fairly accurate, while the size of the peak
is more uncertain. Most importantly, however, the {\em true}
epidemic curve (solid red line) appears to be always enveloped by
prediction curves.

\begin{figure}[tbhp]
\centerline{\includegraphics[height=6cm]{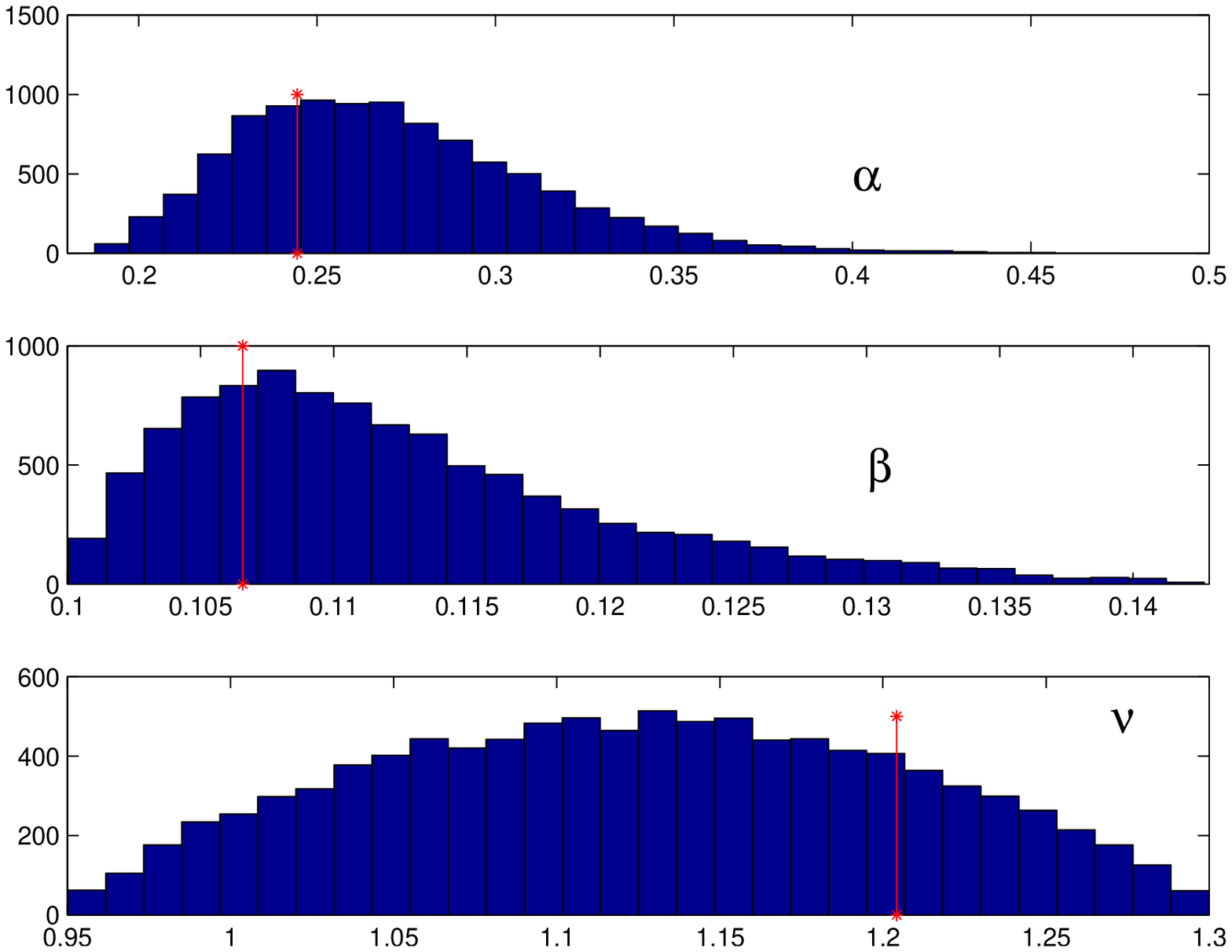}}
\centerline{(a)}
\centerline{\includegraphics[height=6.5cm]{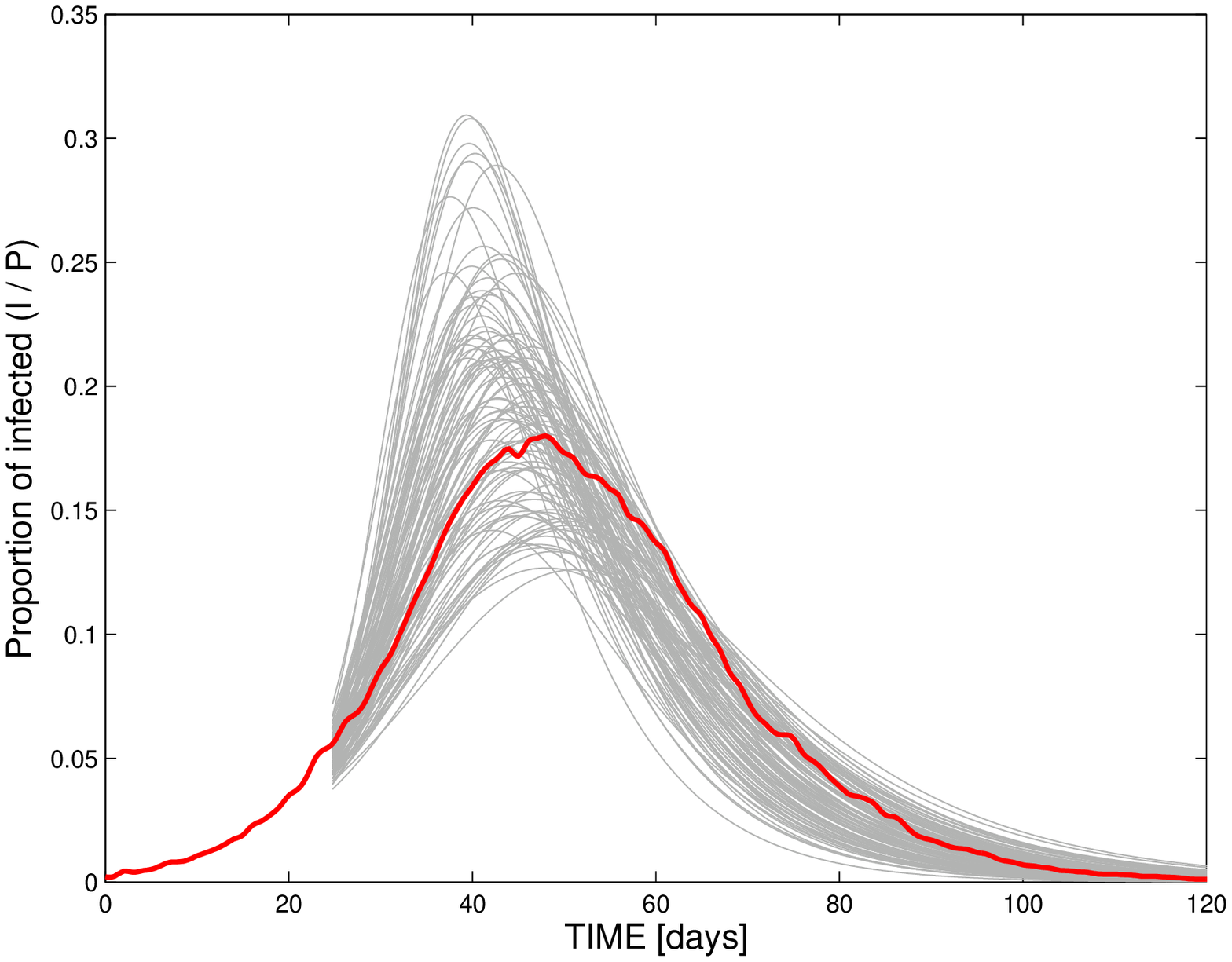}}
\centerline{(b)}
 \caption{Estimation/prediction results from the particle filter after processing measurements collected
 over $25$ days of
surveillance assuming that $\varsigma\in[0.85,1.15]$ (true value
$\varsigma=1.05$): (a) the histograms of estimated parameters
$\alpha$, $\beta$ and $\nu$ (true values indicated by vertical red
lines); (b) Prediction results for a random sample of $100$
particles (gray lines); the red line is the experimental curve from
Fig.\ref{f:exper}}
 \label{f:2}
\end{figure}

The second run of the proposed simulation setup corresponds to the
case with fairly imprecise knowledge of $\varsigma$,  that is
$\varsigma\in[0.85, 1.15]$. The results are shown in Fig. \ref{f:2}.
Comparing Figs. \ref{f:1} and \ref{f:2} one can make two
observations: first, in accordance with our expectations, the
estimation and prediction results are more uncertain when knowledge
of $\varsigma$ is imprecise; second, even when knowledge of
$\varsigma$ is imprecise, the curve corresponding to the {\em true}
number of infected people (solid red line) is always enveloped by
predictions.

\section{Conclusions}

We have developed an algorithm for syndromic surveillance of an epidemic outbreak
formulated in the context of stochastic nonlinear filtering.  The
dynamics of the epidemic is modeled using a compartmental epidemiological model with inhomogeneous mixing which explicitly includes a parameter responsible for the level of social interactions. This model enables simulation of a rich variety of social behavior that may be observed in a small community in response to epidemic (i.e. fear, self-isolations, panic).

The syndromic (typically non-medical) observations  are used  as an algorithm input (e.g. web quries, Twitter messages, sales of certain products, absenteeism from work/study etc). The algorithm (a
particle filter) provides
continuous estimation of the state of the epidemic, including the number of infected people
and the unknown parameters of the model. The numerical results indicate that the proposed
framework can provide useful early prediction of the epidemic peak
if the uncertainty in prior knowledge of model parameters (including social behavior) is not
excessive.

Our future work would encompass a generalization of the proposed
framework to include detection and prediction of other social and
behavioral processes that can be described by  similar
population models (computer worm attacks, propaganda campaigns,
complex social engagements, sensor networks, quorum sensing, etc,
see \cite{Moreno_04}, \cite{Eubank_08}, \cite{Skvortsov_09}).

\section{Acknowledgment}

We thank R.Gailis and C.Woodruff  (DSTO) for valuable discussions
and  E.Yee (DRDC) for drawing our attention to publication
\cite{AMIA2006}.

%\newpage
%\begin{figure}[htb]
%\centerline{\includegraphics[height=6cm,width=7.5cm]{last_fig.eps}}
% \caption{\em RMS position error $\varepsilon_{k}$ for $\alpha=0.1$, $0.5$, $0.99$. }
%\label{f:last}
%\end{figure}
\bibliography{strings}

\begin{thebibliography}{10}

\bibitem{anderson_04}
R.~M. Anderson, C.~Fraser, and A.~C. Ghani.
\newblock Epidemiology, transmission dynamics and control of {SARS}: the
  2002-2003 epidemic.
\newblock {\em Philos. Trans. R. Soc. B Biol. Sci.}, 359:1091--1105, 2004.

\bibitem{anderson_79}
R.~M. Anderson and R.~M. May.
\newblock Population biology of infectious diseases: {P}art 1.
\newblock {\em Nature}, 280:361--367, 1979.

\bibitem{pf_tute}
M.~S. Arulampalam, S.~Maskell, N.~Gordon, and T.~Clapp.
\newblock A tutorial on particle filters for non-linear/non-{G}aussian
  {B}ayesian tracking.
\newblock {\em IEEE Trans. Signal Processing}, 50(2):174--188, Feb. 2002.

\bibitem{Fraser}
C.~Fraser at~al.
\newblock Pandemic potential of a strain of influenza {A (H1N1}): Early
  findings.
\newblock {\em Science}, 324(5934):1557 -- 1561, 2009.

\bibitem{Cauchemez}
S.~Cauchemez and N.~M. Ferguson.
\newblock Likelihood-based estimation of continuous-time epidemic models from
  time-series data: {A}pplication to measles transmission in {L}ondon.
\newblock {\em J. R. Soc. Interface.}, 5(25):885–897, 2008.

\bibitem{Cazelles_97}
B.~Cazelles and N.~P. Chau.
\newblock Using the kalman filter and dynamic models to assess the changing
  {HIV/AIDS} epidemic.
\newblock {\em Mathematical Bioscienses}, 140(2):131--154, 1997.

\bibitem{KnDiscDataMining}
A.~Culotta.
\newblock Detecting influenza outbreaks by analyzing twitter messages.
\newblock In {\em Proc. 2010 Conf. on Knowledge Discovery and Data Mining},
  Washington, D.C., 2010.

\bibitem{Daley_book}
D.J. Daley and J.~Gani.
\newblock {\em Epidemic Modelling}.
\newblock Cambridge Univ. Press, 1996.

\bibitem{Dangerfield}
C.~E. Dangerfield, J.~V. Ross, and M.~J. Keeling.
\newblock Integrating stochasticity and network structure into an epidemic
  model.
\newblock {\em J. R. Soc. Interface}, 6(38):761--774, 2009.

\bibitem{dargatz}
C.~Dargatz.
\newblock A diffusion approximation for an epidemic model.
\newblock Technical report, Ludwig-Maximilian Universitat Munchen, 2007.

\bibitem{Egat_08}
C.~Egat, F.C. Arrat, C.~L. Ajaunie, and H.~Wackernagel.
\newblock Early detection and assessment of epidemics by particle filtering.
\newblock In {\em Proc. 6th Europ. Conf. Geostatistics for Environmental
  Applications {(GeoENV VI)}}, pages 23–--35, Rhodes, Greece, 2008. Springer.

\bibitem{Eubank_08}
S.~Eubank, V.~S.~A. Kumar, and M.Marathe.
\newblock {\em Epidemiology and wireless communication: {T}ight analogy or
  loose metaphor?}
\newblock Springer, 2008.

\bibitem{AMIA2006}
G.~Eysenbach.
\newblock Infodemiology: Tracking flu-related searches on the web for syndromic
  surveillance.
\newblock In {\em Proc. 2006 Symp of American College Med. Inf. Ass. {(AMIA
  2006)}}, pages 244--248, 2006.

\bibitem{Ginsberg_09}
J.~Ginsberg, M.H. Mohebbi, R.~S. Patel1, L.~Brammer, M.~S. Smolinski, and
  L.~Brilliant.
\newblock Detecting influenza epidemics using search engine query data.
\newblock {\em Nature}, 457:1012--1015, 2009.

\bibitem{gordon_etal_93}
N.~J. Gordon, D.~J. Salmond, and A.~F.~M. Smith.
\newblock Novel approach to nonlinear/non-{G}aussian {B}ayesian state
  estimation.
\newblock {\em IEE Proc.-F}, 140(2):107--113, 1993.

\bibitem{jazwinski_70}
A.~H. Jazwinski.
\newblock {\em Stochastic Processes and Filtering Theory}.
\newblock Academic Press, 1970.

\bibitem{jegat_epid}
C.~J\'{e}gat, F.~Carrat, C.~Lajaunie, and H.~Wackernagel.
\newblock Early detection and assessment of epidemics by particle filtering.
\newblock In A.~Soares, M.~J. Pereira, and R.~Dimitrakopoulos, editors, {\em
  geoENV VI – Geostatistics for Environmental Applications}, volume~15, pages
  23--35. Springer, 2008.

\bibitem{kita96b}
G.~Kitagawa.
\newblock Monte {C}arlo filter and smoother for non-{G}aussian non-linear state
  space models.
\newblock {\em Journal Of Computational and Graphical Statistics}, 5(1):1--25,
  1996.

\bibitem{Dailey}
L.Dailey, R.~E. Watkins, and A.~J. Plant.
\newblock Timeliness of data sources used for influenza surveillance.
\newblock {\em J. Am. Medical Inf Ass.}, 14(5):177–185, 2007.

\bibitem{mahler_07}
R.~Mahler.
\newblock {\em Statistical Multisource Multitarget Information Fusion}.
\newblock Artech House, 2007.

\bibitem{Mandela_10}
J.~Mandela, J.~D. Beezleya, L.~Cobba, and A.~Krishnamurthya.
\newblock Data driven computing by the morphing fast {F}ourier transform
  ensemble {K}alman filter in epidemic spread simulations.
\newblock {\em Procedia Computer Science}, 1(1):1221--1229, 2010.

\bibitem{Mnatsakanyan_10}
Z.~R. Mnatsakanyan, H.~S. Burkom, M.~R. Hashemian, and M.~A. Coletta.
\newblock Distributed information fusion models for regional public health
  surveillance.
\newblock {\em Information Fusion}, 12(2):doi:10.1016/j.inffus.2010.12.002,
  2011.

\bibitem{Moreno_04}
Y.~Moreno, M.~Nekovee, and A.~Vespignani.
\newblock Efficiency and reliability of epidemic data dissemination in complex
  networks.
\newblock {\em Phys. Rev. E}, 69(5):055101--1--5, 2004.

\bibitem{Novozhilov}
A.~S. Novozhilov.
\newblock On the spread of epidemics in a closed heterogeneous population.
\newblock {\em Mathematical Bioscienses}, 215(2):177–185, 2008.

\bibitem{ong_10}
J.~B.~S. Ong, M.~I-C. Chen, A.~R. Cook, H.~C. Lee, V.~J. Lee, R.~T.~P. Lin,
  P.~A. Tambyah, and L.~G. Goh.
\newblock Real-time epidemic monitoring and forecasting of {H1N1}-2009 using
  influenza-like illness from general practice and family doctor clinics in
  singapore.
\newblock {\em PLoS ONE}, 5(4):e10036, 2010.

\bibitem{OUDJANEMUSSO}
N.~Oudjane and C.~Musso.
\newblock Progressive correction for regularized particle filters.
\newblock In {\em Proc. 3rd Int. Conf. Information Fusion}, Paris, France,
  2000.

\bibitem{ristic_11}
B.~Ristic.
\newblock Bayesian estimation with imprecise likelihoods: Random set approach.
\newblock {\em IEEE Signal Processing Letters}, 2011.
\newblock In review.

\bibitem{pfbook}
B.~Ristic, S.~Arulampalam, and N.~Gordon.
\newblock {\em Beyond the {K}alman filter}.
\newblock Artech House, 2004.

\bibitem{icassp09}
B.~Ristic, A.~Skvortsov, and M.~Morelande.
\newblock Predicting the progress and the peak of an epidemic.
\newblock In {\em Proc. IEEE Inter. Conf. Acoustics, Speech and Signal
  Processing (ICASSP 2009)}, pages 513--516, Taipei, Taiwan, April 2009.

\bibitem{Roy}
M.~Roy and M.~Pascual.
\newblock On representing network heterogeneities in the incidence rate of
  simple epidemic models.
\newblock {\em J. Ecological Complexity}, 3(1):80--90, 2006.

\bibitem{Schuster_10}
N.~M. Schuster, M.A. Rogers, and L.F. Jr.
\newblock Using search engine query data to track pharmaceutical utilization:
  {A} study of statin.
\newblock {\em The American Journal of Managed Care}, 16(8):e215--e219, 2010.

\bibitem{modsim07}
A.~Skvortsov, R.~Connell, P.~Dawson, and R.~Gailis.
\newblock Epidemic modelling: Validation of agent-based simulation by using
  simple mathematical models.
\newblock In {\em International Congress on Modelling and Simulation {(MODSIM
  2007)}}, pages 657--662, Christchurch, New Zealand, December 2007.

\bibitem{biocomp_07}
A.~Skvortsov, R.~Connell, P.~Dawson, and R.~Gailis.
\newblock Epidemic spread modeling: Alignment of agent-based simulation with a
  simple mathematical model.
\newblock In {\em Proc. Int. Conf. Bioinformatics \& Comput. Biology}, pages
  487--890, Las Vegas, USA, June 2007. CSREA Press.

\bibitem{Skvortsov_09}
A.~Skvortsov, B.~Ristic, and M.~Morelande.
\newblock Networks of chemical sensors: {A} simple mathematical model for
  optimisation study.
\newblock In {\em Proc. 5th International Conference on Intelligent Sensors,
  Sensor Networks and Information Processing {(ISSNIP 2009)}}, pages 385–--390,
  Melbourne, Australia, 2009.

\bibitem{skvortov_10_fus}
A~Skvortsov, B~Ristic, and C~Woodruff.
\newblock Predicting an epidemic based on syndromic surveillance.
\newblock In {\em Proc. 13th Int. Conf. Information Fusion}, Edinburgh, UK,
  July 2010.

\bibitem{stroud_06}
P.~D. Stroud, S.~J. Sydoriak, J.~M. Riese, J.~P. Smith, S.~M. Mniszewski, and
  P.~R. Romero.
\newblock Semi-empirical power-law scaling of new infection rate to model
  epidemic dynamics with inhomogeneous mixing.
\newblock {\em Mathematical Bioscinese}, 203:301--318, 2006.

\bibitem{Herwaarden}
O.~A. van Herwaarden and J.~Grasman.
\newblock Stochastic epidemics: {M}ajor outbreaks and the duration of the
  endemic period.
\newblock {\em J. Math. Biology}, 33(4):581--601, 1995.

\bibitem{Wagner_06}
M.~Wagner, A.~Moore, and R.~Aryel.
\newblock {\em Handbook of Biosurveillance}.
\newblock Elsevier, 2006.

\bibitem{Walden}
J.~Walden and E.~H. Kaplan.
\newblock Estimating time and size of bioterror attack.
\newblock {\em Emerging Infectious Diseases}, 10(7):1202--1205, July 2004.

\bibitem{Wilson_06}
A.G. Wilson, G.D. Wilson, and D.H. Olwell.
\newblock {\em Statistical Methods in Counterterrorism: Game Theory, Modeling,
  Syndromic Surveillance, and Biometric Authentication}.
\newblock Springer, 2006.

\end{thebibliography}
\end{document}